\documentclass[aps,prl,showpacs,twocolumn]{revtex4}
\usepackage{graphicx}
\usepackage{dcolumn}
\usepackage{bm}

\begin{document}

\title{Coherent control of nanomagnet dynamics via ultrafast spin torque pulses.}

\author{Samir Garzon$^{1}$, Longfei Ye$^{1}$, Richard A. Webb$^{1}$,
Thomas M. Crawford$^{1}$, Mark Covington$^{2}$ \& Shehzaad Kaka$^2$}

\affiliation{
 $^1$ Physics and Astronomy Department and USC Nanocenter,
 University of South Carolina, Columbia, SC 29208, USA.
 $^2$ Seagate Research, 1251 Waterfront Place, Pittsburgh, Pennsylvania 15222, USA.
}

%\begin{abstract}
%\end{abstract}
\pacs{72.25.Ba,73.63.-b,75.75.+a}

 \maketitle

\textbf{The magnetization orientation of a nanoscale ferromagnet can
be manipulated using an electric current via the spin transfer
effect~\cite{Slonczewski:1996JMMM,Slonczewski:1999JMMM,Berger:1996PRB,berger:2156JAP}:
spin angular momentum is transferred from the conduction to the
localized electrons, exerting an effective torque on the ferromagnet
~\cite{Katine:PRL2000,Sun:JMMM1999,Tsoi:PRL1998,Myers:Science1999}.
Time domain measurements of nanopillar devices at low temperatures
have directly shown that magnetization dynamics and reversal occur
coherently over a timescale of
nanoseconds~\cite{krivorotov:2005science}. By adjusting the shape of
a spin torque waveform over a timescale comparable to the free
precession period (100-400 ps), control of the magnetization
dynamics in nanopillar devices should be
possible~\cite{rivkin:2006apl,thomas:2006nature,thomas:2007science}.
Here we report coherent control of the free layer magnetization in
nanopillar devices using a pair of current pulses as narrow as 30 ps
with adjustable amplitudes and delay. We show that the switching
probability can be tuned over a broad range by timing the current
pulses with the underlying free-precession orbits, and that the
magnetization evolution remains coherent for more than 1 ns even at
room temperature. Furthermore, we can selectively induce transitions
along free-precession orbits and thereby manipulate the free
magnetic moment motion. In contrast with previous measurements where
the spin torque is applied throughout a large fraction of a
precession
cycle\cite{tulapurkar:2004apl,kaka:2005JMMM,Schneider:2007APL,
devolder:2007prb,devolder:2008PRL}, in our experiments the
magnetization evolves through free-precession except for short time
intervals when it is driven by the spin torque. We expect this
technique will be adopted for further elucidating the dynamics and
dissipation processes in nanomagnets, and will provide an
alternative for spin torque driven spintronic devices, such as
resonantly pumping microwave
oscillators~\cite{Rippard:PRL2004,Kaka:Nature2005}, and ultimately,
for efficient reversal of magnetic memory bits in nanoscale magnetic
random access memory (MRAM).}

We study spin transfer nanopillar devices patterned into $\sim$100nm
ellipses with different aspect ratios (inset of Fig.~1a) at room
temperature and 77 K. Antiferromagnetic dipolar field coupling
between the thick layer (polarizer) and the ``free'' layer is
canceled by biasing the devices with an easy axis magnetic field
$H_\|\sim $800 Oe. The ``free'' layer can be switched between low
resistance (parallel, $P$) and high resistance (anti-parallel, $AP$)
states via a spin-transfer torque from an applied dc current. A
simple shaped waveform, consisting of two current pulses with equal
polarity, comparable amplitude, and separated by a time delay $t_D$
(inset of Fig.~1b), is used to induce and control nanomagnet
dynamics, while the final state of the multilayer is probed by
measuring its steady state resistance. In our devices the free
precession period $\tau\sim$300 ps is much larger than the current
pulse width $\tau_w\sim$30 ps FWHM, but comparable to the
inter-pulse delay 0 ns $< t_D <$ 2 ns. A femtosecond mode-locked
laser in single-shot mode is used to generate a pair of optical
pulses which are converted to electrical pulses using a LT-GaAs/Au
photoconductive switch~\cite{Auston:1984,Smith:APL1989}. A 40 GHz
bias tee is used to inject both the current pulses that induce
magnetization dynamics and the ac/dc currents used to measure the
resistance and reset the device. Reflection measurements show that
typical room temperature pulsewidths at the device are $\sim$30 ps,
but due to cryostat bandwidth limitations the typical pulsewidths
are $\sim$58 ps at 77 K. At nonzero temperatures thermal excitation
of the ``free'' layer magnetization ($\vec{M}$) affects its
dynamical evolution. However, reproducibility in nanomagnet
switching can be increased by applying transverse
fields~\cite{devolder:2007prb} or through inter-layer
coupling~\cite{krivorotov:2005science}. Throughout our measurements
we apply in-plane transverse fields $H_{\perp}\sim$200 Oe to shift
the stable points of $\vec{M}$ away from the easy axis (inset of
Fig.~1a).

\begin{figure}
\begin{center}
\includegraphics[width=3.4in]{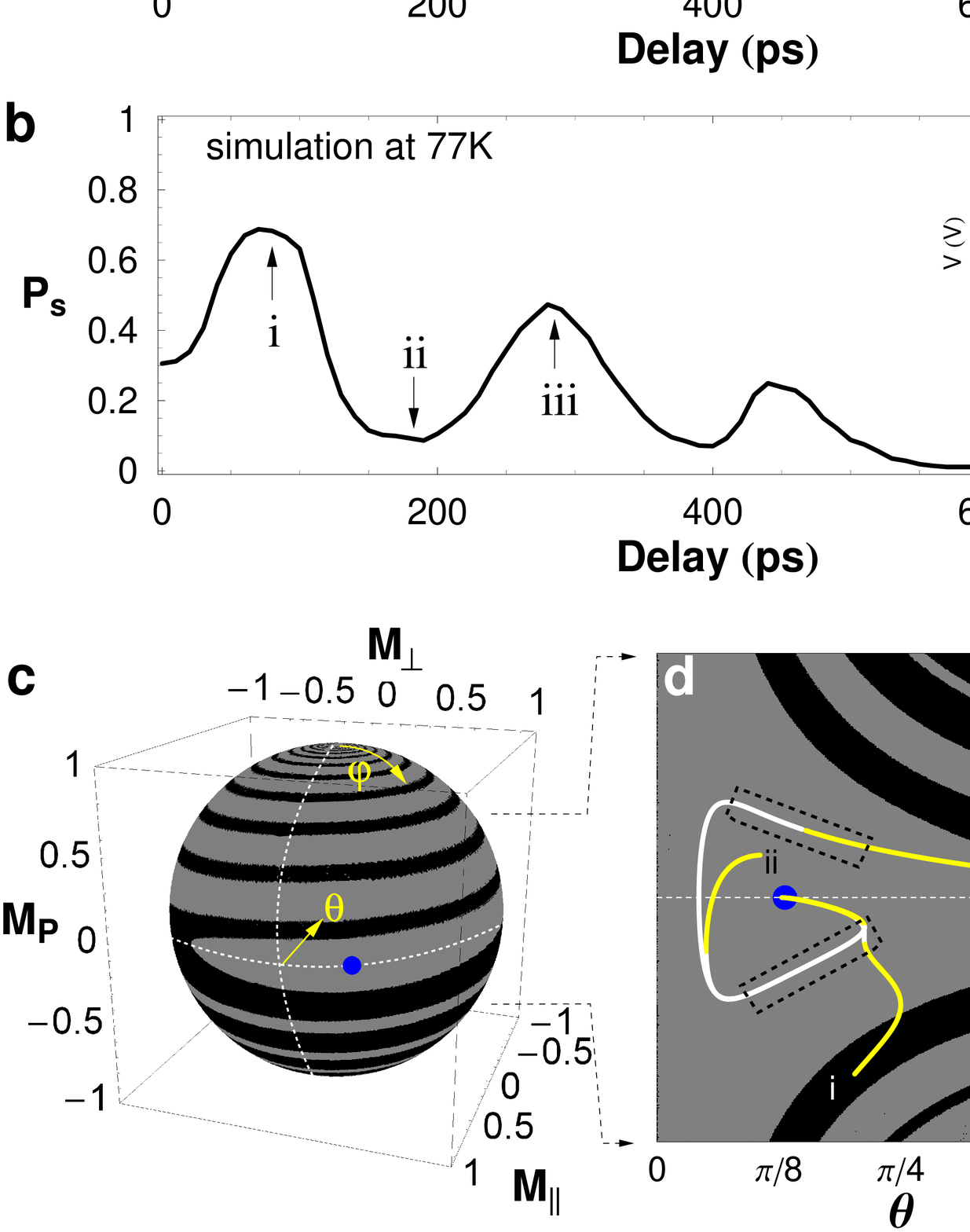}
\caption{\textbf{Modulation of switching probability with delay.}
\textbf{a,} $P_S$ of device N2 at 77 K as a function of delay
between two $\sim$22 mA current pulses. \textbf{Inset,} Schematic of
a Co$_{90}$Fe$_{10}$(8.7nm)/Cu(3nm)/Co$_{90}$Fe$_{10}$(2nm) (type
``N'') nanopillar device. [Ni$_{80}$Fe$_{20}$(20nm)/
Co$_{90}$Fe$_{10}$(2nm)]/Cu(10nm)/Co$_{90}$Fe$_{10}$(3nm) (type
``E'') devices have an extended bottom layer [NiFe/CoFe] to decrease
magnetic layer dipolar coupling. A transverse field is used to shift
the parallel ($P$) and anti-parallel ($AP$) fixed points (blue and
red dot respectively) away from the easy axis. \textbf{b,}
Simulation of the delay dependence of the switching probability due
to two 58 ps FWHM pulses, for an elliptical free layer of size
125nm$\times$75nm$\times$2nm with saturation magnetization $M_S$=800
emu/cm$^3$, a Stoner-Wohlfarth switching field $H_k$=200 Oe, and a
transverse field $H_{\perp}=$80 Oe at 77 K. The labeled regions
correspond to the orbits shown in Fig.~1d. \textbf{Inset,} Pair of
$\sim$30 ps pulses with a delay $t_D=$700 ps measured at a pick-off
tee before the device. \textbf{c,} Phase portrait of $\vec{M}$
showing the basins of attraction for the two stable points $P$
(blue) and $AP$ (red, not visible). Initial conditions
$\theta$,$\varphi$ within the gray (black) basin lead to
no-switching (switching). $\theta$ is the polar angle measured from
the nanomagnet easy axis and $\varphi$ is the azimuthal angle
measured from the normal to the nanomagnet plane shown in the inset
of Fig.~1a. \textbf{d,} $\vec{M}$ trajectories generated by two
current pulses of equal amplitude that have been delayed by 90 ps
(i), 190 ps (ii), and 280 ps (iii). The rectangles enclose regions
where a second pulse has high probability of switching
$\vec{M}$.}\label{fig:fig1}
\end{center}
\end{figure}

The switching probability $P_S$ as a function of inter-pulse delay
for equal amplitude pulses at 77 K is shown in Fig.~1a. Large
modulation of $P_S$ with delay implies coherent dynamics, since
incoherent dynamics would lead to a delay-independent switching
probability $P_2=1-(1-P_1)^2$ (with $P_1$ the single-pulse switching
probability). To understand the origin of these oscillations, we
model the time evolution of the magnetization of a single domain
nanomagnet driven by a perpendicular spin current~(inset of
Fig.~1a), by using a modified Landau-Lifshitz-Gilbert equation which
includes the Slonczewski spin torque
term~\cite{Slonczewski:1999JMMM,sun:2000PRB}. We assume that the
magnetization of the polarizer is fixed, and consider the effect of
nonzero temperatures only on the distribution of initial conditions
but not on the evolution of $\vec{M}$, which is assumed to be
completely deterministic. $\vec{M}$ is described by the angles
$\theta$ and $\varphi$ (inset of Fig.~1a)\cite{sun:2000PRB}, and has
fixed points at $\varphi=\pi/2$, $\theta=\arcsin{H_{\perp}/H_k}$
with $H_k$ the easy axis anisotropy field. The phase portrait of
$\vec{M}$ in the absence of spin torque is shown in
Fig.~1c\cite{Yar:PRB2004}. The black and gray regions, which are the
basins of attraction for the red and blue minimum energy points, are
wrapped around each other, emphasizing the final state's large
sensitivity to fluctuations in initial conditions (i.e. thermal
effects).

Simulations of the delay dependence of $P_S$ for a pair of pulses
with equal amplitude at 77 K (Fig.~1b) show oscillations with delay
that agree qualitatively with our observations (Fig.~1a). Typical
trajectories at consecutive maxima and minima of $P_S$, regions
labeled i, ii, and iii, in Fig.~1b are shown in Fig.1d, where the
section $3\pi/8<\varphi<5\pi/8$ of the phase portrait shown in
Fig.~1c has been stretched into a plane. The initial condition and
first pulse (in yellow) are equivalent for all trajectories, but the
second pulse (also in yellow) is applied at different times
($t_D=$90 ps, 190 ps, and 280 ps). The free evolution between the
two pulses is shown in white. We observe that there are two regions
(dashed boxes in Fig.~1d) where the second spin torque pulse can
more effectively induce basin boundary crossing and lead to
magnetization reversal. As indicated by trajectory ii, a second
pulse applied outside of the marked regions can even push $\vec{M}$
closer to the blue fixed point, effectively cancelling the effect of
the first pulse. Therefore, when a pulse with a width larger than
the free precession period is used for magnetization switching,
partial cancellation of the spin torque can occur, decreasing the
switching probability.

We observe that the clear oscillations and strong modulation of
$P_S$ present in device N2 at 77 K (Fig.~1a) disappear at room
temperature (Fig.~2a). Type ``N'' devices, with dc switching
currents $\sim$0.4 mA, have a small stability factor
$\Delta=E_B/k_BT$ (with $E_B$ the energy barrier between $P$ and
$AP$ states), and thus are more sensitive to thermal effects. At
room temperature such devices typically show decay in $P_S$ with
increasing delay, and small amplitude $P_S$ oscillations. On the
other hand, extended bottom layer (type ``E'') devices, with
switching currents $\sim$2 mA and therefore higher thermal
stability, typically show clear large amplitude oscillations in
$P_S$ both at 77 K (Fig.~2b) and room temperature (Fig.~2c). Fourier
analysis of the oscillations of device E1 at 77 K shows a
fundamental period of 120 ps ($\omega$=8.3GHz) and a much smaller
2$\omega$ harmonic. Since the precession period is twice the period
of the $P_S$ oscillations, for type ``E'' devices $\tau\sim$240 ps.
At room temperature the switching probability of device E2 can be
tuned between 4$\%$ and 93$\%$ by only adjusting the delay between
pulses. The enhancement in the switching probability from 60$\%$ at
zero delay (single pulse) to $\sim$94$\%$ at 120 ps delay (Fig.~2c)
has been measured while keeping the amplitude of the pulses
constant. However, if the total energy delivered by the pulses is
kept constant, a more dramatic enhancement in $P_S$ from 10$\%$ to
70$\%$ at intermediate pulse amplitudes and from 40$\%$ to 95$\%$ at
larger pulse amplitudes is observed. Therefore, multiple current
pulses timed with the underlying coherent dynamics require less
total energy than a single pulse to reproducibly switch spin
transfer devices.

\begin{figure}
\begin{center}
\includegraphics[width=3.4in]{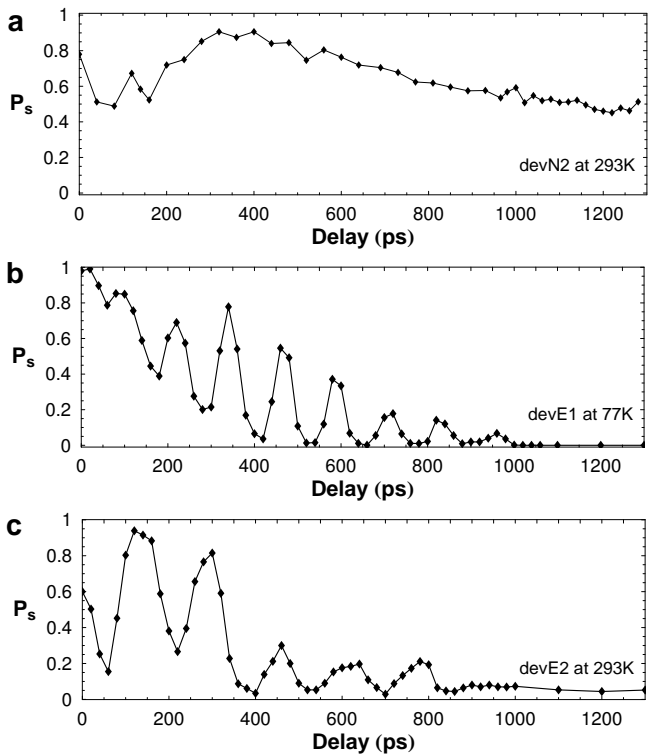}
\caption{\textbf{Effect of temperature on switching probability
modulation.} $P_S$ as a function of delay for: \textbf{a,} device N2
at room temperature. \textbf{b,} device E1 (with extended bottom
layer) at 77 K. \textbf{c,} device E2 (with extended bottom layer)
at room temperature. Type ``E'' devices require larger switching
currents than type ``N'' devices due to increased thermal stability.
Therefore, large modulation of $P_S$ can be observed in typical type
``E'' devices even at room temperature.}\label{fig:fig2}
\end{center}
\end{figure}

We also measured the switching probability as a function of the
amplitude of a pair of pulses while keeping their delay (185 ps) and
relative amplitude ($I_1/I_2=1$) constant (Fig.~3a). $P_S$ initially
increases with increasing pulse amplitude, but after 15 mA it
decreases from 80$\%$ to 55$\%$ before finally increasing to
$\sim$100$\%$ at 23 mA. This counterintuitive result that increasing
the spin torque leads to a decrease in the switching probability is
fully consistent with coherent precession and is predicted by our
simulations~(Fig.~3b). This agreement demonstrates that in our
system the macro-spin model captures the essence of nanomagnet
dynamics. Typical magnetization trajectories at the three labeled
regions of Fig.~3b are shown in Fig.~3c. As the amplitude of the
pair of pulses is increased from region I to region II (Fig.~3b),
the state of $\vec{M}$ at the end of the second pulse moves from the
black basin to a higher energy gray basin region, therefore
decreasing $P_S$ (Fig.~3c). As the amplitude of the pulses is
increased further to region III in Fig.~3b, the first pulse produces
enough spin torque to switch $\vec{M}$ (Fig.~3c). Thus it is
possible to move the magnetization into larger angle, higher energy
orbits by applying multiple short current pulses with controlled
amplitudes and delays.

\begin{figure}
\begin{center}
\includegraphics[width=3.5in]{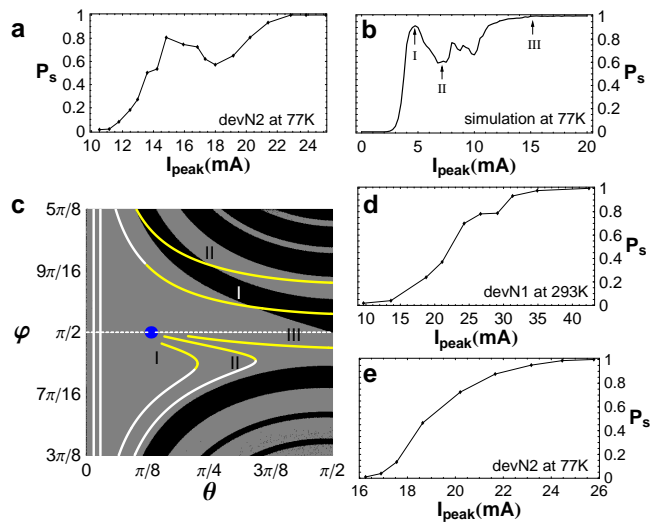}
\caption{\textbf{Amplitude dependence of the switching probability.}
\textbf{a,} $P_S$ at 77 K as a function of the amplitude of a pair
of $\sim$58 ps pulses with a fixed delay of 185 ps, for type ``N''
device 2. The amplitudes of both pulses are kept equal. \textbf{b,}
Simulated switching probability for the situation described in
\textbf{a} for the same parameters as Fig.~1b. \textbf{c,} $\vec{M}$
trajectories at the labeled regions of Fig.~3b corresponding to
pulse amplitudes of 4.8 mA (I), 6.8 mA (II), and 15 mA (III).
Initial conditions have a thermal probability distribution.
\textbf{d,} $P_S$ at room temperature as a function of the amplitude
of a single $\sim$30 ps pulse for type ``N'' device 1. \textbf{e,}
$P_S$ at 77 K as a function of the amplitude of a single $\sim$58 ps
pulse for type ``N'' device 2.}\label{fig:fig3}
\end{center}
\end{figure}

In contrast to previous
reports~\cite{tulapurkar:2004apl,kaka:2005JMMM,Schneider:2007APL}
where single pulses with $\tau_w>$100 ps were required to achieve
large $P_S$, we demonstrate $P_S\sim$100$\%$ with single 30 ps
pulses at room temperature (Figs.~3d,e). Furthermore, for devices
with dc switching currents comparable to those previously
reported~\cite{kaka:2005JMMM,Schneider:2007APL} we achieve
$P_S\approx$100$\%$ with pulse amplitudes two times smaller than
expected from the assumption of pulsewidth and amplitude being
inversely proportional~\cite{koch:2004prl,Li:PRB2003}. These results
are supported by macrospin simulations which indicate that the
pulsewidth-current product required for $P_S=$95$\%$ decreases by
more than a factor of two when $\tau_w\ll\tau$. Therefore,
ultrashort current pulses apply spin torque more efficiently,
increasing the probability of magnetization switching. Depending on
field bias, temperature, and device anisotropy, $P_S$ can show
either stepped\cite{devolder:2007prb} (Fig. 3d) or
smooth\cite{Schneider:2007APL,kaka:2005JMMM} (Fig. 3e), but always
monotonic increase with increasing pulse amplitude. The stepped
increase in $P_S$ is predicted by our simulations and was previously
observed when increasing the pulse width~\cite{devolder:2007prb}.
The steps are caused by the underlying free precession orbits, which
play an essential role at short timescales, where the switching
process is driven, instead of thermally-assisted. Thus, the free
precession orbits provide a map for tailoring the amplitudes and
timing of a series of short pulses in order to control the
magnetization evolution.

As long as the motion remains coherent until the arrival of the last
pulse, a series of short current pulses can be used to manipulate
the free layer into a desired free-precession orbit, that is, to
coherently control the magnetization. Our technique not only
provides a proof of principle for such pulsed magnetization control,
but also demonstrates coherent magnetic moment dynamics even at room
temperature, and provides an alternative for high probability low
power device switching. Using current pulses much shorter than the
free-precession period is critical for controlling the magnetization
motion. Our technique can be used to study the switching process in
magnetic tunnel junctions, where a quiet ``incubation'' period that
precedes magnetization switching has been
observed~\cite{devolder:2008PRL}, as well as to probe coherence and
damping in out-of-plane nanopillar devices. Potential applications
of our approach also include
nano-oscillators~\cite{Rippard:PRL2004,Kaka:Nature2005} which could
be resonantly pumped for generating tunable microwaves over a broad
GHz range.

We acknowledge helpful discussions with Yaroslaw Bazaliy. This work
was funded by Seagate Research.

\bibliography{references}

\end{document}